\documentclass[aps,prl,twocolumn,superscriptaddress,showpacs,amsmath,amssymb]{revtex4-1}
\usepackage[utf8]{inputenc}
\usepackage{amsmath,amsthm}
\usepackage{graphicx}
\usepackage[colorlinks = true,
            linkcolor = blue,
            urlcolor  = blue,
            citecolor = blue,
            anchorcolor = blue]{hyperref}
\usepackage{bbold}
\usepackage{braket}
\usepackage{xcolor}
\usepackage[title]{appendix}
\renewcommand{\L}{\mathbf{L}}

\newcommand{\tr}[1]{\ensuremath{\operatorname{tr}\!{#1}}}

\newcommand{\be}{\begin{equation}}
\newcommand{\ee}{\end{equation}}
\newcommand{\bea}{\begin{eqnarray}}
\newcommand{\eea}{\end{eqnarray}}

\newcommand{\one}{\mathbb{1}}

\global\long\def\ga{\gamma} 
\global\long\def\De{\Delta} \global\long\def\Ga{\Gamma}
\global\long\def\th{\theta}
\global\long\def\th{\theta}

\global\long\def\ra{\rightarrow}

\global\long\def\si{\sigma}

\global\long\def\al{\alpha}
\global\long\def\ga{\gamma}

\theoremstyle{thm@}

\theoremstyle{remark}

\begin{document}

\title{
Exact nonequilibrium steady state of open XXZ/XYZ spin-1/2 chain with
Dirichlet boundary conditions}
\author{Vladislav Popkov}
\affiliation{Faculty of Mathematics and Physics, University of Ljubljana, Jadranska 19, SI-1000 Ljubljana, Slovenia}
\affiliation{Bergisches Universit\"at Wuppertal, Gauss Str. 20, D-42097 Wuppertal, Germany}
\author{Toma\v z Prosen}
\affiliation{Faculty of Mathematics and Physics, University of Ljubljana, Jadranska 19, SI-1000 Ljubljana, Slovenia}
\author{Lenart Zadnik}
\affiliation{Faculty of Mathematics and Physics, University of Ljubljana, Jadranska 19, SI-1000 Ljubljana, Slovenia}

\begin{abstract}
We investigate a dissipatively driven XYZ spin-$1/2$ chain in the Zeno limit of strong dissipation, described by Lindblad 
master equation. The nonequilibrium steady state is expressed in terms of a matrix product ansatz using novel site-dependent Lax operators. The components of 
Lax operators  satisfy a simple set of linear recurrence equations that generalize the defining algebraic relations of the quantum 
group ${\cal U}_q(sl_2)$. We reveal connection between the nonequilibrium steady state of the nonunitary dynamics and the respective
integrable model with edge magnetic fields, described by coherent unitary dynamics.
\end{abstract}
\maketitle

\textbf{Introduction.--}
One of the main current efforts of the condensed matter physics society is to understand quantum states of matter far from equilibrium~\cite{Eisert, Bloch}.
Understandably, simple models with tractable but non-trivial exact solutions are of key importance in this game. The realm of driven dissipative quantum many-body systems~\cite{Diehl}
provides nice and rich examples of such models, capable of displaying genuinely out-of-equilibrium phenomena, for example, novel types of non-equilibrium phase transitions \cite{PP,EP,Fleischhauer,Heyl}.
While exact treatment of the aforementioned class of models is essentially limited to quasi-free situations, it is remarkable that some exact solutions have been found even in the case of strong interactions,
in particular in quantum integrable spin chains with dissipation and incoherent driving localized at the chain's boundaries~\cite{JPAreview}. Despite the fact that the exact matrix product form of these solutions has been found only for a very specific choice of the boundary jump operators~\cite{TP2011,KPS2013}, this provided a fresh perspective on the effect of local and quasilocal
conservation laws on quantum transport and relaxation~\cite{QLreview}.
It has, however, remained an open question of how these exact steady state solutions fit into the general framework of integrability. For example, except in the special case of  dissipatively driven noninteracting models~\cite{fabian}, the solvable dissipatively driven boundaries cannot be generated using the solutions of the ubiquitous reflection equations~\cite{sklyanin}, which constitute the standard framework for generating integrable boundaries in the coherent (nondissipative,
Hamiltonian) setting.

In the present letter we make a step forward in the understanding of integrability of open XYZ spin-$1/2$ chain, dissipatively driven at the boundaries. We present 
an exact solution for the nonequilibrium steady state of the model with arbitrary boundary processes, as long as the edge spins are described by pure states in the large dissipation limit. The matrix product ansatz for the steady state is inhomogeneous and drastically different from the previously known exact solutions that describe steady states of the spin chains driven along a particular axis at both edges.
Firstly, we introduce the model and describe how the  solution is constructed by unveiling an intriguing relationship between two systems: a coherent quantum system on the one hand, and open spin chain with strong boundary dissipation, described by a Lindblad master equation
for the density matrix, on the other hand. The results are presented in the second part of the Letter. We discuss, in particular, the appearance of sharp resonances (``ballistic" windows) in the anisotropy dependence of the spin current in the special case of the XXZ model. Lastly we outline the algebraic content of the site-dependent Lax operators constituting the inhomogeneous matrix product ansatz for the steady state. The mathematical details needed to fully grasp our formal constructions can be found in the acompanying publication~\cite{2019PPZ-PRE}.

\begin{figure}[ht!]
\centering
\hspace{-0.25cm}\includegraphics[width=1\linewidth]{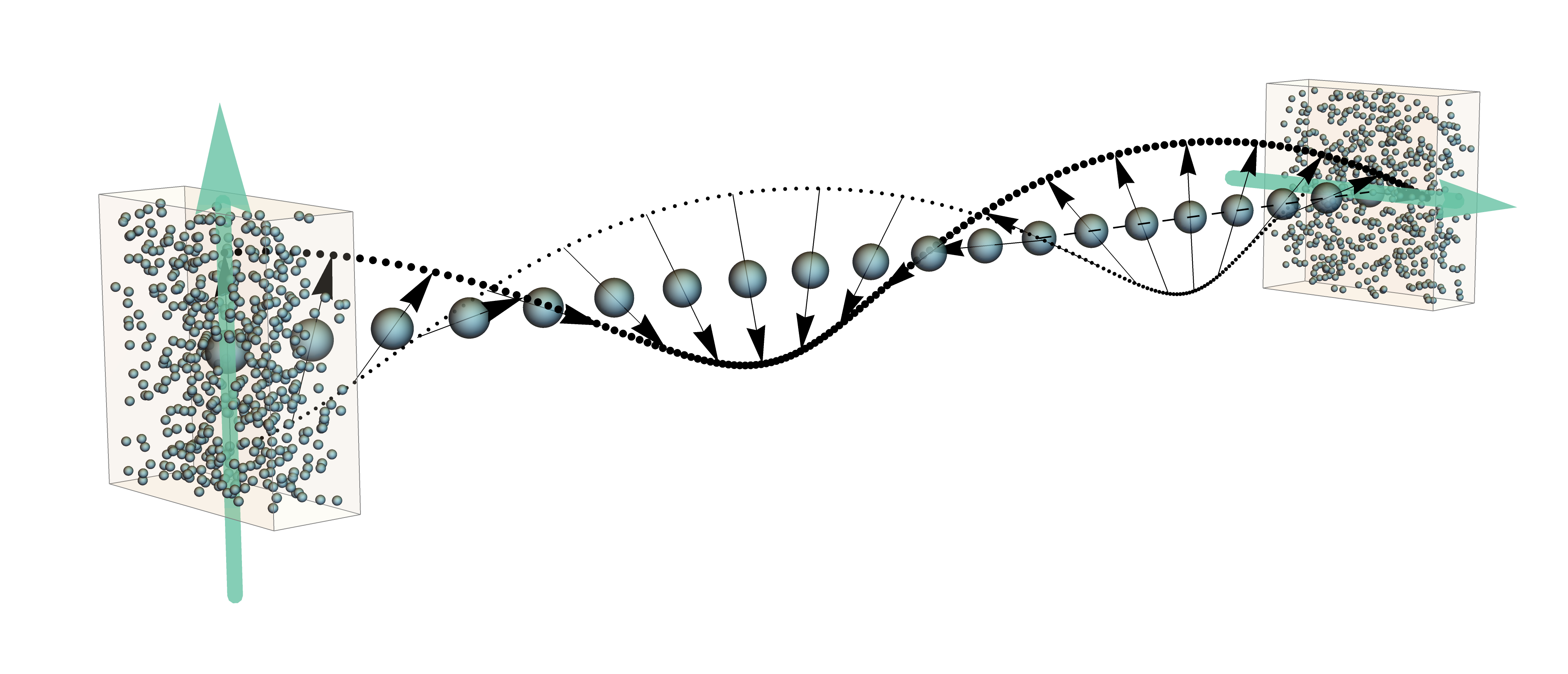}
\caption{Schematic setup: spin chain, which is attached to two fully polarizing reservoirs at its boundaries. The bulk follows a coherent evolution, while the edge spins are, in addition, driven by dissipative polarising processes, resulting from the coupling to reservoirs.}
\label{fig:setup}
\end{figure}

\textbf{The model.--} The setting under investigation is depicted in Fig.~\ref{fig:setup}.
We wish to construct the nonequilibrium steady state (NESS) of the Lindblad equation
\be
\frac{{\rm d}}{{\rm d}t} \rho(t) = -i [H,\rho(t)] +\Gamma\,{\cal D}_{\rm l}[\rho(t)]+\Gamma\,{\cal D}_{\rm r}[\rho(t)],
\label{eq:LME}
\ee
at large dissipation strength $\Gamma$, where the dissipators ${\cal D}_{\mu}[\rho]=2k^{}_{\mu}\rho k^\dagger_{\mu}-\{k^\dagger_{\mu}k^{}_{\mu},\rho\}$, for $\mu\in\{{\rm l},{\rm r}\}$, act at the left and right ends of the chain of $N+2$ sites, labelled by $0$ and $N+1$, respectively. The local jump operators are $k^{}_{\rm l,r}=(\vec{n}'_{\rm l,r} + i \vec{n}''_{\rm l,r})\cdot \vec{\sigma}_{0,N+1}$, so that the dissipators target polarizations $\vec{n}_{\mu}=\vec{n}(\theta_{\mu},\phi_{\mu})$, where $\vec{n}(\theta,\phi)=(\sin \theta \cos \phi,\sin \theta\sin \phi,\cos \theta)$. Here, real vectors $\vec{n}'_{\mu} = \vec{n}(\frac{\pi}{2}-\theta_{\mu},\pi+\phi_{\mu})$, $\vec{n}''_{\mu} = \vec{n}(\frac{\pi}{2},\phi_{\mu}-\frac{\pi}{2})$ and $\vec{n}_{\mu}$ form an orthonormal basis of $\mathbb R^3$.
The targeted states of the dissipators are single-site pure states $\rho_{\mu}$, such that ${\cal D}_{\mu}[\rho_{\mu}]=0$ and $\tr[\rho_\mu\vec{\sigma}]=\vec{n}_\mu$. The unitary part of the evolution is generated by an XYZ spin-$1/2$ Hamiltonian
\be
H = \sum_{n=0}^{N}  \vec{\sigma}_n \cdot {\rm J} \vec{\sigma}_{n+1},
\label{eq:Hamiltonian}
\ee
in which ${\rm J} = {\rm diag}(J_x,J_y,J_z)$ denotes the anisotropy tensor, $\sigma^\alpha_n$ for $\alpha\in \{x,y,z\}$ are Pauli matrices and we have denoted $\vec{\sigma}_n = (\sigma^x_n,\sigma^y_n,\sigma^z_n)$.

The general spatially homogeneous XYZ model, in which the dissipation rates on both edges are equal, has eight free parameters, which we collect into a vector $\boldsymbol{\Pi}=(N,J_y/J_x,J_z/J_x,\phi_{\rm l},\th_{\rm l},\phi_{\rm r},\th_{\rm r},\Ga/J_x)$.
We are interested in the steady state solution of the Lindblad equation~\eqref{eq:LME}, denoted by
\be
\rho_{\infty}(\boldsymbol{\Pi})= \lim_{t \rightarrow \infty} \rho(\boldsymbol{\Pi},t).
\label{NESS}
\ee
The first fully analytic solution of a model of interacting spins (XXZ model with $J_z/J_x=J_z/J_y=\De$)
was proposed in Ref.~\cite{TP2011}
for arbitrary dissipation strength $\Ga$ and boundary polarizations along the $z$-axis:
$\phi_{\rm l}=\phi_{\rm r}=\th_{\rm l}=0$, $\th_{\rm r}=\pi$.
The steady state~\eqref{NESS} was calculated exactly via a matrix product ansatz (MPA)
\begin{align}
 &\rho_{\infty}(\boldsymbol{\Pi}) =\Omega\,\Omega^\dagger,\qquad\Omega=\bra{0}\L_0\ldots \L_{N+1}\ket{0},
\label{eq:MPA-1}
\end{align}
in which each Lax operator $\L_n$ is a $2\times 2$ matrix in the physical space $\L_n\equiv\sum_{\al\in\{x,y,z\}} L^\al \si_n^\al $
with elements $L^\al$ acting in the auxiliary space. Operators $L^x\pm i L^y$ and $L^z$ form
a non-unitary highest-weight representation of the ${\cal U}_q(sl_2)$ algebra, where 
$\tfrac{1}{2}(q+q^{-1})=\De$ parametrizes the anisotropy and $\ket{0}$ is the highest-weight state of the representation~\cite{KPS2013}. 
Since Lax operators $\L_n$, used in Eq.~\eqref{eq:MPA-1}, differ only in the physical site upon which they act (denoted by $n$), 
we call the ansatz~\eqref{eq:MPA-1} a homogeneous matrix product ansatz (MPA). Such a construction was used in all previously exactly solved models~\cite{JPAreview}. In contrast, in this Letter we introduce a fundamentally different inhomogeneous MPA with site-dependent Lax matrices [see, for example, Eq.~\eqref{eq:matrixels}], to solve a general problem~\eqref{NESS}, with arbitrary polarization angles $\phi_\mu,\theta_\mu$, in the limit $\Gamma\to\infty$.

In Ref.~\cite{JPAreview} it was recognized that, for arbitrary value of $\Ga$ and $|\De| \neq 1$, the specific boundary conditions $\phi_{\rm l}=\phi_{\rm r}=\th_{\rm l}=0$, $\th_{\rm r}=\pi$ determine the only possible dissipative setup allowing for analytic treatment with the proposed homogeneous MPA method~\cite{TP2011}.
It has, however, become clear that with growing system size $N$ the steady state becomes
independent of $\Ga$, i.e. an effective quantum Zeno regime~\cite{SudarshanZeno1977,KoshinoPhysRepZeno2015} is reached (in the isotropic XXX case this happens for $N \gg 1/\Ga$; see~\cite{PSK2013-PRE}). Later, the general tendency of locally-dissipative one-dimensional quantum models to approach strong dissipation regime for growing $N$ was established on the basis of Lieb-Robinson bound~\cite{Znidaric2015}. The formal Zeno limit $\Ga \rightarrow \infty$ typically corresponds to the thermodynamic limit $N\gg 1$ for arbitrarily small $\Ga$ and vice-versa: for any finite $N$ one has $\rho_{\infty}(\boldsymbol{\Pi}) \sim \rho_{\infty}^{\rm Zeno}$ for some $\Ga \gg \Ga_{\rm char}(N,\ldots)$, where $\Ga_{\rm char}$ is some characteristic dissipation strength and $\ldots$ denote other parameters.

In this Letter we show, how to derive the full analytic NESS~\eqref{NESS} for
an arbitrary choice of anisotropy constants $J_\al$ and arbitrary boundary angles
$\phi_\mu$, $\th_\mu$ in the Zeno limit $\Ga \ra \infty$. Put differently, we resolve the generally posed problem of computing the NESS~\eqref{NESS}, apart from the assumption of the Zeno-regime which can be viewed as a thermodynamic regime $N \gg 1$ of the model. As such, our solution opens the possibility of analyzing its thermodynamic features.
In particular, in the special case $\boldsymbol{\Pi}=(N,J_y/J_x=1,J_z/J_x=\Delta,\phi_{\rm l},\th_{\rm l},\phi_{\rm r},\th_{\rm r},\Ga)$, we provide an explicit form of
\be
\lim_{\Ga \ra \infty} \lim_{t \ra \infty}
\rho(\boldsymbol{\Pi},t)=\lim_{\Ga\to\infty}\rho_\infty(\boldsymbol{\Pi})= \rho_{\infty}^{\rm Zeno}
\label{NESS-Zeno}
\ee
[see Eq.~\eqref{eq:MPA-2} and Eq.~\eqref{eq:matrixels} below],
the uniqueness of which is guaranteed, for any choice of the parameters, by the Evans criterium~\cite{Evans}. The most straightforward application of our results would be construction of the steady state phase diagram of the open dissipative XYZ spin chain with fixed and pure boundary states $\rho_{\rm l}$ and $\rho_{\rm r}$, which would be analogous to solving an initial value Dirichlet-type quantum problem. Obviously, the universal features of the phase diagram can become transparent only in the thermodynamic limit that enforces the Zeno regime, and can be attained, at least in principle, within our approach. Previous studies~\cite{TP2011} could not answer this question since the only boundary-related parameter was the amplitude of the dissipation.

\textbf{Zeno limit solution.--}
In the Zeno regime $\Ga\gg 1$ the states at the edges of the spin chain are fixed after the time of order $1/\Ga$~
\cite{Venuti,2018PopkovZenoDynamics} so that
$\rho_{\infty}^{\rm Zeno}
=\rho_{\rm l} \otimes R \otimes\rho_{\rm r}$, where $\rho_{\rm l}$ and $\rho_{\rm r}$ are the targeted single-spin pure states and $R$ denotes the bulk of the steady state. For the latter we postulate
\begin{align}
&R =\Omega\Omega^\dagger,\qquad\Omega=\bra{0}\L_1\ldots \L_{N}\ket{\psi},
\label{eq:MPA-2}
\end{align}
where  $\ket{\psi}$ and $\ket{0}$ are some vectors in the auxiliary space, that shall be specified below by requiring the compatibility conditions~\eqref{BEleft} and~\eqref{BEright}.
Formally, the ansatz~\eqref{eq:MPA-2} has a form similar to~\eqref{eq:MPA-1}, with the number of local Lax matrices reduced by two, due to factoring out of the two boundary spins, which is a result of the strong dissipation. After the fast relaxation of the boundary spins the bulk follows an effective coherent evolution~\cite{Venuti} with subleading slow dissipative relaxation towards the steady state. The unitary part of the evolution is generated by the dissipation-projected Hamiltonian $H_{\cal D}$, which commutes with the bulk part of the steady state: $[R,H_{\cal D}]=0$. In our case it has the same form as the Hamiltonian of the initial XYZ model, truncated to internal sites $1,\ldots N$,
with additional local magnetic fields at sites $1$ and $N$~\cite{2018PopkovZenoDynamics,Venuti}:
\be
H_{\cal D}= \sum_{n=1}^{N-1}  h_{n,n+1}+ \left( {\rm J} \vec{n}_{\rm l} \right)\vec{\si}_1+
\left( {\rm J} \vec{n}_{\rm r} \right)\vec{\si}_N,
\label{eq:hD}
\ee
where $h_{n,n+1}=\vec{\sigma}_n \cdot {\rm J} \vec{\sigma}_{n+1} $ is the local density of the Hamiltonian.
Notably, the magnetic fields are determined by the anisotropy and the direction of the targeted polarization.

We now postulate Lax operators to be of the form $\L_{n}=\sum_{\al\in\{x,y,z\}}L_n^\al \si_n^\al$ and denote by $I_n$ unit operators in physical (quantum) and auxiliary spaces. The commutativity $[R,H_{\cal D}]=0$ then follows from the local divergence condition $[h_{n,n+1},\L_n \L_{n+1}]=
i I_n \L_{n+1}-i\L_{n} I_{n+1}$, which holds for all $n=1,\ldots N-1$, provided that boundary equations (analogues of the reflection equation~\cite{sklyanin})
\begin{align}
\bra{0} [2({\rm J} \vec{n}_{\rm l})\cdot\vec{L}_{1} + i\,I_1]&=0,\label{BEleft}\\
[2({\rm J} \vec{n}_{\rm r})\cdot\vec{L}_{N} - i\,I_N]\ket{\psi}&=0
\label{BEright}
\end{align}
are satisfied. Remarkably, the divergence condition can be rewritten as a pair of {\em discrete Landau-Lifshitz} equations
\be
\vec{L}_n \times {\rm J} \vec{L}_{n+1} = \frac{1}{2} \vec{L}_n I_{n+1},\quad
{\rm J}\vec{L}_n \times \vec{L}_{n+1} = \frac{1}{2} I_n \vec{L}_{n+1}
\label{eq:recurrence}
\ee
for the vectors $\vec{L}_n=(L^x_n,L^y_n,L^z_n)$. In the XXZ model, these equations define an extension of the $\mathcal{U}_q(sl_2)$ algebra with an infinite number of generators; see Ref.~\cite{2019PPZ-PRE}. In this case we find an explicit solution for their elements $L^\alpha_n$, while in the general XYZ case we can generate them numerically via recurrence~\eqref{eq:recurrence}.

Surprisingly,  Lax matrices $\L_n$ appear to differ element- and shape-wise from site to site. The dimension of the auxiliary
space depends on the lattice site index $n$ so that the components $L_n^\al$  are
rectangular matrices of size $n \times (n+1)$ with $n$-dependent elements -- product $\bra{0}L_1^{\al_1}\ldots L_{N}^{\al_{N}}$ is thus a vector in $\mathbb C^{N+1}$. The unit matrices are simply $I_n=\sum_{k=0}^{n-1} \ket{k}\bra{k}\one$, while
for the chosen gauge~\eqref{eq:MPA-2} Lax matrices depend also on the left boundary
conditions, i.e. the angles of polarization $\phi_{\rm l}$, $\th_{\rm l}$. We remark that, given $\vec{L}_{n}$, recurrence~\eqref{eq:recurrence} is an overdetermined set of equations for $\vec{L}_{n+1}$, so the very existence of a solution is exceptional, hinting at hidden symmetries and possible integrability.

\textbf{The XXZ case.--}
For the uniaxial XXZ case, where
$J_x=J_y=J, \ \ J_z/J=\De$, we rewrite Lax matrices satisfying~\eqref{eq:recurrence} as $\L_n=
L_n^{+} \si_n^{+} +  L_n^{-} \si_n^{-}+  L_n^{z} \si_n^{z}$. The components in this basis read (see \cite{2019PPZ-PRE}  for a proof):
\begin{align}
\begin{gathered}
L_n^z\!=\!\sum_{k=1}^n \ket{k}\!\bra{k+1},\, \,L_n^\pm=\!\pm\eta^{\mp 1}\!\sum_{k=0}^{n-1} \sum_{l=0}^{n}\De^n
B_{n;k,l}^{\pm} \ket{k}\!\bra{l}, \\
B_{n;k,l}^{\pm}\!=\!\left[\frac{\pm i}{2 \De} \right]^{k-l+1}\!
\sum_{m=0}^{l}\!\binom{n-k-1}{m}\!\binom{n-m}{l-m} [i\De^{-1}]^{2m},
\label{eq:matrixels}
\end{gathered}
\end{align}
where $\eta$ is a free parameter. Equations~\eqref{BEleft} and~\eqref{BEright} provide the dependence of NESS on the boundary conditions. 
Indeed, substitution of $\vec{L}_{1}$ into the left-boundary equation~\eqref{BEleft} fixes 
 $\eta= -e^{i \phi_{\rm l}} \tan(\th_{\rm l}/2)$ in all $L_{n}^\pm$.
On the other hand, the right boundary conditions enter the MPA~\eqref{eq:MPA-2} through the boundary vector $\ket{\psi}\in {\mathbb C}^{N+1}$
satisfying~\eqref{BEright}. For specific cases $\ket{\psi}$ takes simple forms, for instance, for $\De=0$ one has $\ket{\psi}=\ket{N}$,  independently of boundary polarizations.
For $\th_{\rm l}=\th_{\rm r}=\pi/2$, and  $\phi_{\rm r}-\phi_{\rm l}=(N+1)\ga$, where $\De=\cos \ga $, we find $\langle k | {\psi}\rangle=(-2 \sin \ga)^{N-k}$, corresponding to spin helix states~\cite{2016PopkovPresilla}. We note that the right boundary equation~\eqref{BEright} has at least one nontrivial solution for $\ket{\psi}$, which is unique for a generic choice of boundary polarizations. In these cases $R$ from Eq.~\eqref{eq:MPA-2} is uniquely defined.

With $R$ fixed, the  operator $\rho_{\infty}^{\rm Zeno}=\rho_{\rm l}\otimes R\otimes\rho_{\rm r}$ correctly reproduces the NESS of the Lindblad equation (\ref{eq:LME}) in the Zeno limit.
In special cases it agrees perfectly with the known analytical solutions~\cite{2012XYtwist}, while in generic case, comparison with numerically exact NESS, computed as prescribed in~\cite{2018PopkovZenoDynamics}, yields equivalence up to numerical machine precision. Also, for finite values of the dissipation strength $\Gamma$, the ansatz~\eqref{eq:MPA-2} for $\rho_{\infty}^{\rm Zeno}$ converges towards the solution of $i[H,\rho(\Gamma)]=\Gamma\,{\cal D}_{\rm l}[\rho(\Gamma)]+\Gamma\,{\cal D}_{\rm r}[\rho(\Gamma)]$, so that in the operator norm $\|\rho_{\infty}^{\rm Zeno}-\rho(\Gamma)\|=O(\Ga^{-1})$ for large $\Ga$ (data not shown). This again indicates that the ansatz~\eqref{eq:MPA-2} is correct. Note that there are also cases, in which the right boundary vector
$\ket{\psi}$ in the MPA is not unique. This typically happens when extra degeneracy is present in the spectrum of the dissipation-projected Hamiltonian~\eqref{eq:hD}. It only appears for fine-tuned parameters, so we conjecure the degeneracy of $\ket{\psi}$ to happen on a measure-zero subset of the full parameter space. Even on this zero-measure submanifold, though, the Zeno NESS is correctly reproduced by our ansatz~\eqref{eq:MPA-2}, if the right boundary vector is chosen correctly.

\begin{figure}[ht!]
\centering
\includegraphics[trim={0 0 0 0},clip,width=1\linewidth]{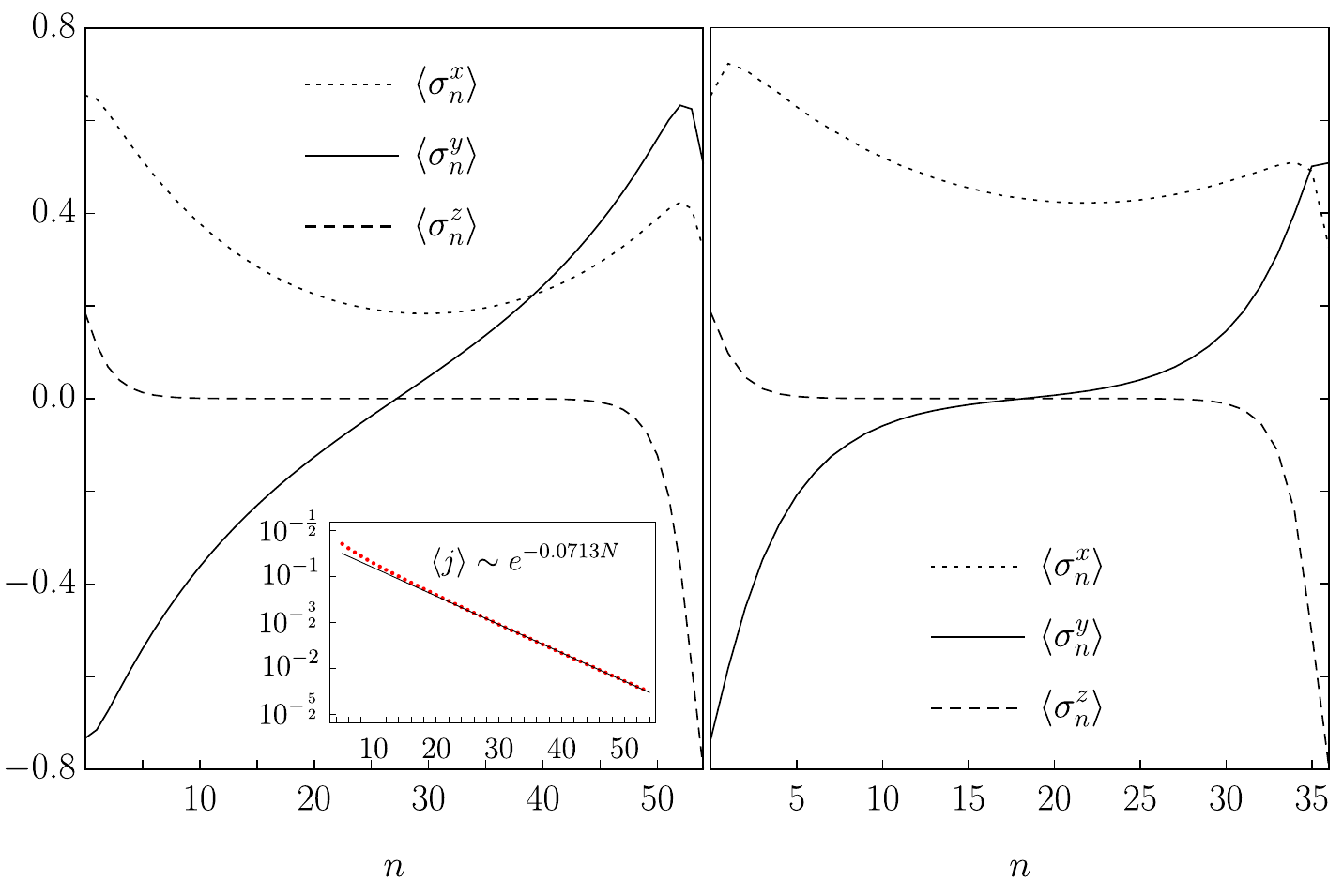}
\caption{Profiles of magnetization in XXZ spin chain (left) and XYZ spin chain (right). 
The inset on the left graph shows exponential decay of the current with system size in the XXZ case.
This is a generic example of our problem, parameters being $\phi_{\rm l}=\sqrt{3}\pi$, $\theta_{\rm l}=(1-\sqrt{5}/4)\pi$, $\phi_{\rm r}=\sqrt{5}\pi/7$ 
and $\theta_{\rm r}=(7-\sqrt{5})\pi/6$. In the XXZ case $\gamma=(\sqrt{5}-1)\pi/8$ and in XYZ case $J_x=13/10$, $J_y=6/5$, $J_z=1$.
 System sizes (without the sites on which the jump operators act) are $N=53$ and $N=35$, respectively.}
\label{fig:profiles_scaling}
\end{figure}

\textbf{The XYZ case.--} Fully anisotropic model with all $J_\al$ different is treated analogously, 
using a generalized $1\times 2$ seed for $L_1^\al$; for details see~\cite{2019PPZ-PRE}.
Here, higher $L_n^\al$ are not known analytically and are generated by numerically solving the recurrence~\eqref{eq:recurrence}.
Like in the XXZ case, one can efficiently calculate numerically exact Zeno NESS for large system sizes; see, for instance, magnetization profiles in the right panel of Fig.~\ref{fig:profiles_scaling}.

\begin{figure}[ht!]
\centering
\includegraphics[trim={5 0 0 0},clip,width=1\linewidth]{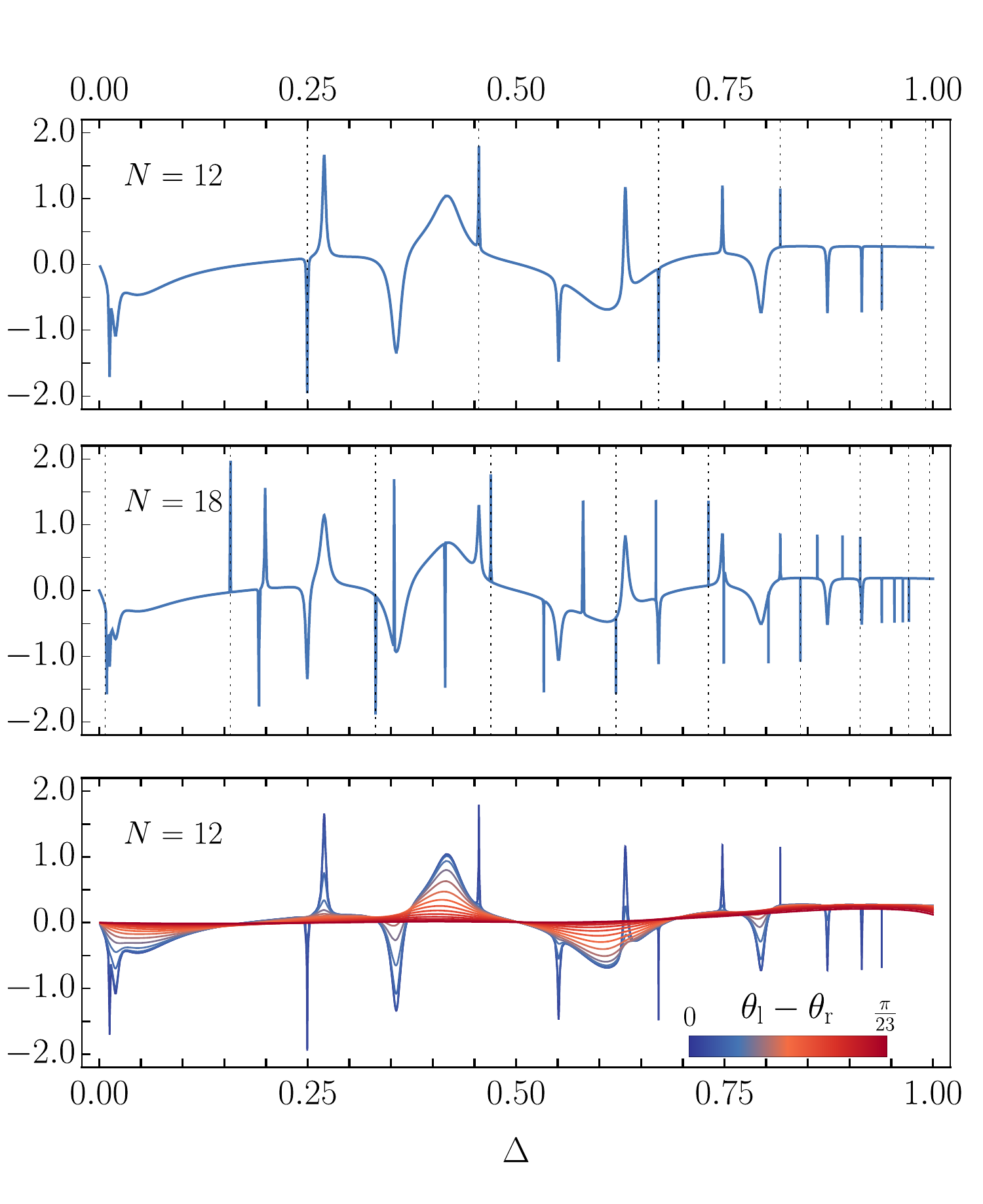}
\caption{Steady-state magnetization current $j^z=2 \langle \si_n^x \si_{n+1}^y -
\si_n^y \si_{n+1}^x \rangle$ versus anisotropy $\De=J_z/J_x=J_z/J_y$ in a driven XXZ spin chain, for $N=12,18$. Two upper frames
refer to the $xy$-plane boundary gradient  $\theta_{\rm l}=\theta_{\rm r}=\pi/2$,
$\phi_{\rm r}-\phi_{\rm l}=25\pi/46$.
Resonances indicated with dotted lines
correspond to the spin helix states \cite{2016PopkovPresilla}.
In the lower frame $j^z$ is shown for gradually
increasing polar angle mismatch
$0<\th_{\rm l}-\th_{\rm r}<\pi/23$, while $\th_{\rm r}=\pi/2$ is kept constant.}
\label{fig:currents}
\end{figure}

\textbf{Observables in the NESS.--}
The MPA~\eqref{eq:MPA-2} allows for an efficient computation of local one- and two-point
observables, for instance magnetization profiles, spin current $j^z$ in the XXZ model, and others, for previously inaccessible
system sizes; see, e.g. Fig.~\ref{fig:profiles_scaling}.
A particularly interesting phenomenon is plotted in Fig.~\ref{fig:currents}, where the spin current $j^z$ in
the XXZ model is shown to exhibit a high sensitivity with resonance spiking as a function
of the anisotropy parameter $J_z/J_x=\De$. With increasing system size $N$ the resonances sharpen and become more dense, while
their positions change. In the resonances the current is approximately ballistic $j^z=O\left( 1  \right)$, while it decreases with $N$ on the slowly-varying background, on top of which the resonances are formed.

Such sensitivity of the boundary driven XXZ chain with respect to the anisotropy can be traced to the existence of the
current-carrying eigenstates in the spectrum of the dissipation projected Hamiltonian $H_{\cal D}$ [given in~\eqref{eq:hD}]
for matching polar boundary angles $\th_{\rm l}=\th_{\rm r}\equiv \th$. In particular, we can identify
the spin-helix states (SHS) among the spikes. They are characterized by ballistic magnetization current and helical
magnetization profile~\cite{2016PopkovPresilla,2017PopkovPresillaJohannesJPA} and appear at
critical values of the anisotropy: $\De_{\rm cr}=\cos \ga_m$, $\ga_m=(2 \pi m + \phi_{\rm r} - \phi_{\rm l})/(N+1)$. 
The magnetization current of the SHS depends on the helicity angle $\gamma_m$, but not on the system size $N$, 
namely $j^z=2 (\sin \th)^2 \sin \gamma_m$.
Remarkably, the ``SHS'' anisotropies $\De_{\rm cr}$ are the only points where the Zeno NESS
is a pure state; for all other values of $\De$ it is mixed. In the thermodynamic limit the collection of all resonant points 
forms a dense countable set. We stress again that matching polar angles $\th_{\rm l}\approx\th_{\rm r}$ are the
crucial requirement for observation of these resonances. If $\th_{\rm l}$ and $\th_{\rm r}$ become
significantly different,  the current-carrying states in the spectrum of $H_{\cal D}$ disappear, and so do the resonances; see the lower frame of Fig.~\ref{fig:currents}.

\textbf{Discussion.--}
We have presented an exact steady state solution of a Lindblad problem that describes a spin chain whose boundary spins are collapsed
in arbitrary pure states with pre-fixed magnetization vectors. On the practical level this allows to investigate
setups with up to $50$-$100$ spins, i.e. gives a means to investigate the full phase diagram of an open XYZ spin chain.
We stress that this regime is not accessible to variational MPS methods (aka DMRG), since higher orders in $1/\Gamma$, of the NESS density operator, have volume-law operator entanglement.

From the mathematical point of view we have proposed an inhomogeneous Lax structure that presents a possible alternative construction
of nontrivial commuting operators, not directly related to the Sklyanin's construction based on the reflection algebra~\cite{sklyanin}.    
All Hamiltonians of the type~\eqref{eq:hD} are known to have a generating function for the conservation laws, i.e. they stem from a spectral-parameter dependent transfer-matrix~\cite{OffDiagonal}. Consequently, coherent systems~\eqref{eq:hD} are potentially integrable with one of the versions of the  Quantum Inverse Scattering Method; see e.g.~\cite{takhtajan} for a review on the algebraic Bethe ansatz approach. This fact gives an intriguing link between integrable coherent and dissipatively driven 1D spin chains, at large dissipation strengths.

At the end we should note, that the investigated Lindblad dynamics is realizable with the protocol of repeated interactions~\cite{Karevski2014}, based on the usual Heisenberg (von Neumann) coherent evolution. In short, the boundary atomic magnetic moments have to interact repeatedly with a magnet, freshly polarized in the required direction. The individual addressing of single atomic magnetic moments, utilised in the protocol, is possible with existing experimental techniques~\cite{Kai2019}.

\vspace{2mm}
We acknowledge discussions with M. Petkov\v sek and V. Romanovsky. LZ would like to thank Katja Klobas for
useful comments on the manuscript. The work has been supported by European Research Council (ERC) through
the advanced grant 694544 -- OMNES and the grant P1-0402 of Slovenian Research Agency (ARRS).
V.P. also acknowledges support by the DFG grant KL 645/20-1.


\begin{thebibliography}{99}

\bibitem{Eisert} J.~Eisert, M.~Friesdorf, and C.~Gogolin, \href{https://www.nature.com/articles/nphys3215}{{\em Nature Phys.} {\bf 11}, 124 (2015)}.

\bibitem{Bloch} C.~Gross and I.~Bloch, \href{https://science.sciencemag.org/content/357/6355/995.full}{{\em Science}  {\bf 357}, 995 (2017).}

\bibitem{Diehl} S.~Diehl,  A.~Micheli, A.~Kantian, B.~Kraus, H.~P.~B\" uchler and P.~Zoller, \href{https://www.nature.com/articles/nphys1073}{{\em Nature Phys.} {\bf 4}, 878 (2008).}

\bibitem{PP} T.~Prosen and I.~Pi\v zorn, \href{https://journals.aps.org/prl/abstract/10.1103/PhysRevLett.101.105701}{{\em Phys. Rev. Lett.} {\bf 101}, 105701 (2008).}

\bibitem{EP} J.~Eisert and T.~Prosen, \href{https://arxiv.org/abs/1012.5013}{{\tt arXiv:1012.5013}.}

\bibitem{Fleischhauer} M.~H\" oning, M.~Moos, M.~Fleischhauer, \href{https://journals.aps.org/pra/abstract/10.1103/PhysRevA.86.013606}{{\em Phys. Rev. A} {\bf 86}, 013606 (2012).}

\bibitem{Heyl} M.~Heyl, \href{https://iopscience.iop.org/article/10.1088/1361-6633/aaaf9a/meta}{{\em Rep. Prog. Phys.} {\bf 81}, 054001 (2018).}

\bibitem{JPAreview} T.~Prosen, \href{https://iopscience.iop.org/article/10.1088/1751-8113/48/37/373001/meta}{{\em J. Phys. A: Math. \& Theor.} {\bf 48}, 373001 (2015)}.

\bibitem{TP2011} T. Prosen, \href{https://doi.org/10.1103/PhysRevLett.106.217206}{{\em Phys. Rev. Lett.} {\bf 106}, 217206 (2011)}; \href{https://journals.aps.org/prl/abstract/10.1103/PhysRevLett.107.137201}{{\em Phys. Rev. Lett.} {\bf 107}, 137201 (2011)}.

\bibitem{KPS2013} D.~Karevski, V.~Popkov, G.~M.~Sch\" utz, \href{https://journals.aps.org/prl/abstract/10.1103/PhysRevLett.110.047201}{{\em Phys. Rev. Lett.} {\bf 110}, 047201 (2013)}.

\bibitem{QLreview} E. Ilievski, M. Medenjak, T. Prosen, L. Zadnik, \href{https://doi.org/10.1088/1742-5468/2016/06/064008}{{\em J. Stat. Mech.} ({\bf 2016}) P064008}.

\bibitem{fabian} M.~V.~Medvedyeva, F.~H.~L.~Essler,~T.~Prosen, \href{https://doi.org/10.1103/PhysRevLett.117.137202}{{\em Phys. Rev. Lett.} {\bf 117}, 137202 (2016).}





\bibitem{sklyanin} E.~K.~Sklyanin, \href{https://iopscience.iop.org/article/10.1088/0305-4470/21/10/015}{{\em J. Phys. A: Math. \& Gen.} {\bf 21}, 2375 (1988).}

\bibitem{2019PPZ-PRE} V.~Popkov, T.~Prosen, L.~Zadnik, Joint publication to this work, submitted to Phys. Rev. E.

 \bibitem{SudarshanZeno1977}
B. Misra and E.C.G. Sudarshan,
\textit{J. Math. Phys.} \textbf{18} (1977) 756.

\bibitem{KoshinoPhysRepZeno2015} K. Koshino and A. Shimizu,
\textit{Phys. Rep.} \textbf{412},(2005) 191.

\bibitem{PSK2013-PRE}  V.~Popkov, D.~Karevski and G.~M.~Sch\" utz, \href{https://journals.aps.org/pre/abstract/10.1103/PhysRevE.88.062118}{{\em Phys. Rev. E} {\bf 88}, 062118 (2013)}.

\bibitem{Znidaric2015} \v Znidari\v c M.,
\href{https://journals.aps.org/pre/abstract/10.1103/PhysRevE.92.042143},
 Phys. Rev. E \textbf{92}, (2015) 042143.

\bibitem{Evans} D.E. Evans, {{\em Comm. Math. Phys.}, {\bf 54}, 293 (1977).}

\bibitem{Venuti}
P.~Zanardi, L.~C.~Venuti,
\href{https://doi.org/10.1103/PhysRevLett.113.240406}{{\em Phys. Rev. Lett.} {\bf 113}, 240406 (2014).}

\bibitem{2018PopkovZenoDynamics}
V.~Popkov, S.~Essink, C.~Presilla, G.~M.~Sch\"utz,
\href{https://doi.org/10.1103/PhysRevA.98.052110}{{\em Phys. Rev. A} {\bf 98}, 052110 (2018).}

\bibitem {2016PopkovPresilla}
V.~Popkov, C.~Presilla,
  \textit{Phys. Rev. A} \textbf{93} 022111 (2016)



\bibitem{2012XYtwist}
V.~Popkov, \href{https://doi.org/10.1088/1742-5468/2012/12/P12015}{{\em J. Stat. Mech.} {\bf 2012}, P12015 (2012).}

\bibitem{2017PopkovPresillaJohannesJPA} V. Popkov, C.~Presilla, and J. Schmidt,
\href{https://doi.org/10.1088/1751-8121/aa86cb}{\textit{J. Phys. A} \textbf{50} 2017, 435302 ( 2017).}




\bibitem{OffDiagonal}
 Wang, Y., Yang, W.-L., Cao, J. and  Shi, K.,
 Off-Diagonal Bethe Ansatz for Exactly Solvable Models, Springer, ISBN  978-3-662-46756-5.

\bibitem{takhtajan} L.~D.~Faddeev, L.~A.~Takhtajan, \href{https://doi.org/10.1070/RM1979v034n05ABEH003909}{{\em Uspekhi Mat. Nauk} {\bf 34}:5, 13 (1979).}


\bibitem{Karevski2014} G. T. Landi, E. Novais, M. J. de Oliveira, D. Karevski,
Phys. Rev. E \textbf{90} (2014) 042142.

\bibitem{Kai2019} Kai Yang et al.,
\textit{Phys. Rev. Lett.} \textbf{122} 227203 (2019).







%





%

%
%



\end{thebibliography}
\end{document}